\documentclass[conference]{IEEEtran}
\IEEEoverridecommandlockouts
\usepackage{cite}
\usepackage{amsmath,amssymb,amsfonts}
\usepackage{algorithmic}
\usepackage{graphicx}
\usepackage{textcomp}
\usepackage{xcolor}

\makeatletter
\newcommand{\linebreakand}{%
  \end{@IEEEauthorhalign}
  \hfill\mbox{}\par
  \mbox{}\hfill\begin{@IEEEauthorhalign}
}
\makeatother

\def\BibTeX{{\rm B\kern-.05em{\sc i\kern-.025em b}\kern-.08em
    T\kern-.1667em\lower.7ex\hbox{E}\kern-.125emX}}
\begin{document}

\title{Exploiting Consistency-Preserving Loss and Perceptual Contrast Stretching to Boost SSL-based Speech Enhancement\\

}

\author{\IEEEauthorblockN{Muhammad Salman Khan}
\IEEEauthorblockA{\textit{Kore University of Enna} \\
Enna, Italy \\
muhammadsalman.khan@unikorestudent.it
}
\and
\IEEEauthorblockN{Moreno La Quatra}
\IEEEauthorblockA{\textit{Kore University of Enna} \\
Enna, Italy \\
moreno.laquatra@unikore.it }
\and
\IEEEauthorblockN{Kuo-Hsuan Hung}
\IEEEauthorblockA{\textit{Academia Sinica} \\
Taipei, China \\
khhung@citi.sinica.edu.tw}
\linebreakand 
\IEEEauthorblockN{Szu-Wei Fu}
\IEEEauthorblockA{\textit{NVIDIA Corporation} \\
Taipei, China \\
szuweif@nvidia.com}
\and
\IEEEauthorblockN{Sabato Marco Siniscalchi}
\IEEEauthorblockA{\textit{Università degli Studi di Palermo} \\
Palermo, Italy \\
sabatomarco.siniscalchi@unipa.it}
\and
\IEEEauthorblockN{Yu Tsao}
\IEEEauthorblockA{
\textit{Academia Sinica} \\
Taipei, China \\
yu.tsao@citi.sinica.edu.tw }
}

\maketitle

\begin{abstract}
 Self-supervised representation learning (SSL) has attained SOTA results on several downstream speech tasks, but SSL-based speech enhancement (SE) solutions still lag behind. To address this issue, we exploit three main ideas: (i) Transformer-based masking generation, (ii) consistency-preserving loss,  and (iii) perceptual contrast stretching (PCS).   In detail, conformer layers, leveraging an attention mechanism, are introduced to effectively model frame-level representations and obtain the Ideal Ratio Mask (IRM) for SE. Moreover, we incorporate consistency in the loss function, which processes the input to account for the inconsistency effects of signal reconstruction from the spectrogram.    Finally, PCS is employed to improve the contrast of input and target features according to perceptual importance.  Evaluated on the VoiceBank-DEMAND task, the proposed solution outperforms previously SSL-based SE solutions when tested on several objective metrics, attaining a SOTA PESQ score of 3.54.
\end{abstract}


\section{Introduction}
\label{sec:intro}
Speech enhancement (SE) is a challenging task that aims to improve speech intelligibility and quality by removing unwanted background noise from the audio channel. SE is often used as a front-end task for telecommunications, hearing aids, and automated speech recognition (ASR) \cite{b1,b2,b3}. Broadly speaking, there exist two main SE categories: discriminative and generative. 
In the former category, a discriminative model is trained to minimize the difference between enhanced and clean speech leveraging either a  masking- and a mapping-based approach. In the masking-based solution, either an ideal binary mask (IBM) or a smoothed ideal ratio mask (IRM) \cite{b4} is estimated using a set of complementary features created from the noisy speech. The noisy features are then subjected to the predicted mask to produce the improved features \cite{b5}. In the second approach, a deep model is used to perform a high-dimension vector-to-vector mapping and is trained in a regression mode \cite{b6,b7}. In both approaches, the difference between the enhanced and reference speech is obtained using a point-wise $L_{p}$-norm distance \cite{b8} or exploiting a perceptual metric \cite{b9}.  
In the second category, the distribution of the clean speech is taken into account to generate the objective function. More specifically, generative solutions aim to match the distribution of the speech signal rather than optimizing a point-wise loss with the goal of generating a more natural-sounding speech. Several works utilizing deep generative models for SE have been proposed in recent years, for example, SEGAN \cite{b10}, CMGAN \cite{b11}, and SCP-GAN \cite{b12}  leverages generative adversarial networks \cite{b13} for implicit density estimation. 
An explicit density estimation is instead put forth, for instance, in \cite {b14}, and \cite{b15},  where a variational autoencoder \cite{b16} has been used. 
Diffusion-based generative models \cite{b7} have also been investigated to address SE, e.g., \cite{b18, b19}; the key idea is to gradually transform clean speech into noise, and training a deep model to invert this process for different noise scales. Deep model architecture for processing both magnitude and phase information have also been proposed, e.g., \cite{b20, b21}.
 
In recent years, we have witnessed a surge in interest in self-supervised learning (SSL) approaches, namely, Wav2vec2.0 \cite{b22}, HuBERT \cite{b23}, and WavLM \cite{b24},  to extract speech latent representations that have then been successfully used in downstream speech tasks, as shown in the SUPERB challenge \cite{b25}. Nonetheless, SSL-based SE results were not as good as those attained in the other speech tasks. In \cite{b26}, the authors argued that the key issue was the lack of fine-grained information in SSL embeddings and tried to address the problem by introducing cross-domain features. 
In \cite{b27} several data-strategies have been tried to improve the system performance. In \cite{b44} they used clean SSL for latent space modeling within the Conditional Variational
Autoencoder (CVAE) framework, ensuring the model fully utilizes existing knowledge without noise interference.
Despite those efforts, SSL-based SE solutions still lag behind state-of-the-art (SOTA) SE methods. 
This work aims to address this issue and close the gap with SOTA SE solutions. 
To this end, we introduce specific architectural choices along with an ad-hoc consistency-preserving loss when deploying our discriminative SSL-based SE solution; furthermore, we also perform a spectral pre-processing step on both the noisy and target speech data before training our models. In particular, as a key architectural choice, we used the convolution-augmented Transformer (Conformer) \cite{b28} to generate the mask needed in the enhancement stage. 
Indeed, Conformer has been proven effective in various speech applications \cite{b28, b29}, including SE \cite{b30, b31}. Here we argue that the conformer-based heads allow for better modeling of the contextual information in the speech signals. To optimize the neural parameters, we also used the idea proposed in \cite{b12} for consistency-preserving training. More specifically, the effect of the iSTFT reconstruction, which causes inconsistencies between signals, is taken into account when computing the loss, e.g., \cite{b32, b12}. Finally, perceptual contrast stretching (PCS)~\cite{b33} was employed as a pre-processing step on both the noisy and target waveforms before training our model in order to better exploit the peculiar sensitivity of the human auditory systems and improve the final speech perceptual quality. In testing, PCS was applied only on the noisy waveform. Experimental evidence on the VoiceBank-DEMAND task shows that our solution improves previously proposed SSL-based SE methods, and it consistently reduces the gap to SOTA SE solutions.

\begin{figure*}[h]
    \centering
    \centerline{\includegraphics[width=\linewidth, height=6.5cm]{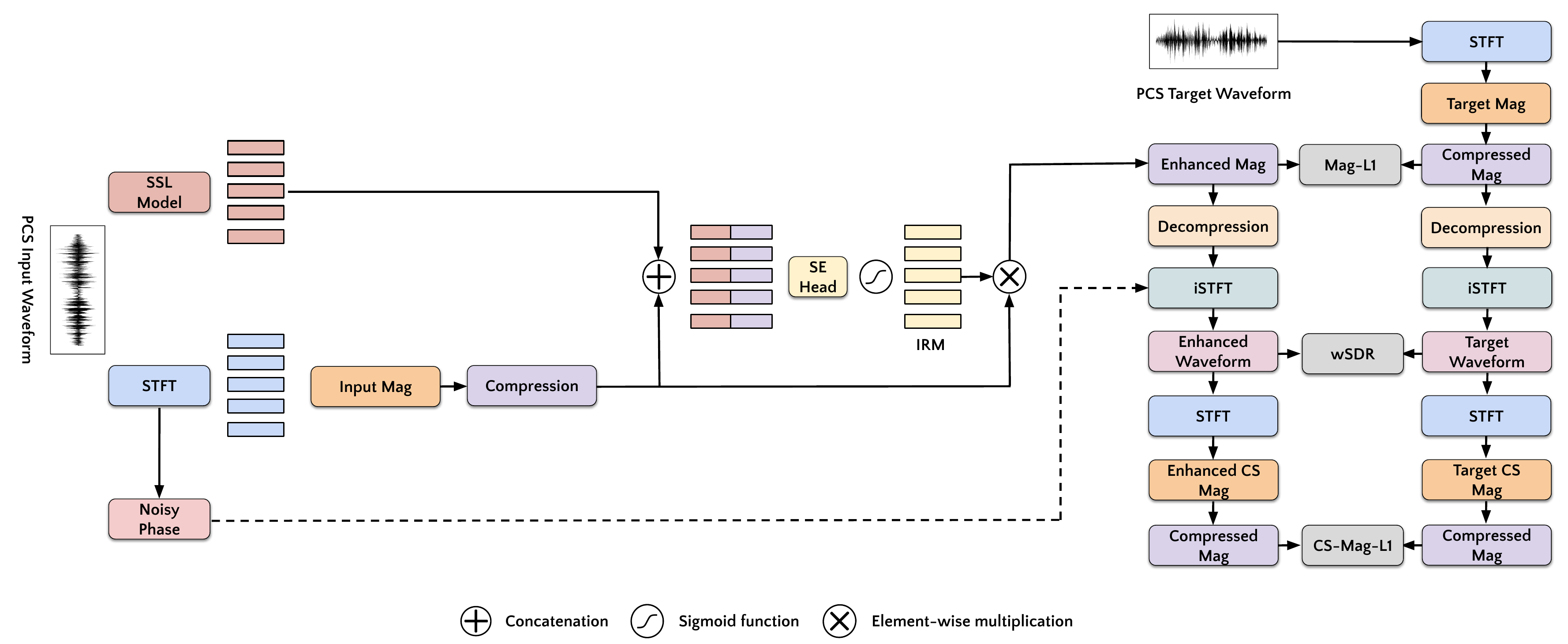}}
    \caption{Overall architecture of the proposed SSL-based speech enhancement (SE) model. The magnitude of the STFT is indicated with Mag. CS stands for consistent, and PCS indicates perceptural contrast stretching.}
    \label{fig:architecture}
\end{figure*}

The rest of the work is organized as follows. The proposed SSL-based SE architecture is given in Section \ref{sec:model}. Section \ref{sec:experiments} presents the dataset used in our experiments. The  results are discussed in Section \ref{sec:results}. Finally, Section \ref{sec:conclusion} concludes our work.
\section{SSL-based speech enhancement model}
\label{sec:model}

\subsection{Overall architecture}
\label{ssec:ssl_block}
The proposed SSL-based SE model consists of two main components, namely: (i) a feature extractor and (ii) a SE head. 
The feature extractor is based on an SSL model that extracts frame-level representations directly from the speech waveform. 
It is initialized using a pre-trained SSL model, whose parameters have been optimized in a self-supervised manner leveraging unlabeled speech data with the goal of learning rich speech representations.
On top of the feature extractor is the SE component, which is a supervised model that takes the speech representations extracted by the feature extractor at its input.  
It outputs a frame-level representation that is used to estimate the ideal ratio mask (IRM) for the input signal.
The two components are integrated into an end-to-end trainable architecture. 

Figure~\ref{fig:architecture} provides a high-level representation of the proposed SE framework. The input speech waveform is first processed by the feature extractor to obtain a frame-level representation using the pre-trained SSL model. Meanwhile, the STFT representation is computed in order to obtain both magnitude and phase spectra. 
Following the approach proposed in \cite{b26}, the frame-level representations extracted by the SSL model and the magnitude spectrum obtained from the STFT are combined to form a unified frame representation that incorporates both the high-level features from the SSL model and the spectral information from the spectrogram, as shown in Figure~\ref{fig:architecture}.
According to previous findings \cite{b26}, the STFT magnitude spectrum is compressed using a logarithmic function (i.e., $log(1 + x)$) before being combined with the SSL features.
As the enhancement process is completed, the enhanced magnitude spectrum is decompressed using an exponential function (i.e., $e^{x} - 1$) to restore its original dynamic range.

\subsection{Speech enhancement head}
\label{ssec:se_block}

The SE head is a supervised model designed to predict  IRM  for the input signal. The IRM represents the ratio between the magnitudes of clean and noisy speech in each time-frequency bin~\cite{b4}.
The component consists of a stack of neural network layers that jointly process the (i) frame-level representations extracted from the pre-trained SSL feature extractor and (ii) compressed magnitude spectrum obtained from the STFT of the input signal.
We investigate different architectural configurations for those neural layers, including Transformer, Conformer, and BiLSTM layers.

All layer types retain the original number of frames when processing the input spectrogram, rather than collapsing it, to exploit temporal dependencies in the input sequence.
For all configurations, the neural network component contains only 2 layers.
The output of the SE head is a frame-level representation used to estimate the IRM via a sigmoid function.
The estimated IRM is multiplied with the noisy magnitude spectrum to obtain the enhanced magnitude spectrum.
The enhanced speech signal is then reconstructed by combining the enhanced magnitude spectrum with the phase from the noisy input - the process is outlined in Figure~\ref{fig:architecture}.
By backpropagating errors through the whole model, comprising the feature extractor and SE component, it can be optimized end-to-end for the speech enhancement task.

\subsection{Loss computation}
\label{ssec:consistency}
In this work we use three losses during the neural parameters optimization stage, namely:
\begin{enumerate}
    \item \textbf{Waveform-based loss (wSDR):} The weighted Signal-to-Distortion Ratio (wSDR)~\cite{b34} loss compares the time-domain waveform of the enhanced speech to the clean reference signal. An advantage of this loss is that is bounded within the range $[-1, +1]$ and also be more phase sensitive, as inverted phase gets penalized as well. Moreover, it compensates for the time delay of the noisy signal, as discussed in~\cite{b35}.
    \item \textbf{Magnitude spectrum L1 loss (mag-L1):} The mean absolute error was used as a loss function and is based on  L1 distance between the compressed magnitude spectra obtained via STFT of the enhanced and clean signals. It aims to encourage alignment in the spectral domain.

    \item \textbf{Consistency-Preserving  Magnitude spectrum L1 loss (CS-mag-L1):} The CS-mag-L1 loss reinforces consistency in magnitude spectrum computation following the methodology outlined in \cite{b12}, where the authors argue that losses used in SE do not take into account the effect of the iSTFT reconstruction which causes inconsistencies between signals. In particular, the authors modified architecture and loss function(s) so that any input going into a loss function undergoes the same process, taking into consideration the effects of signal reconstruction from the spectrogram. We replicated their approach in the right, bottom part of Figure \ref{fig:architecture}.
\end{enumerate}
This comprehensive strategy for loss computation ensures that the model is trained to achieve two key objectives: maintaining fidelity in the waveform representation and ensuring consistency and coherence in the spectral characteristics between the enhanced output and the original input. By optimizing these aspects simultaneously, the model aims to improve the overall quality of the enhanced speech. 
The loss computation process is depicted on the right side of Figure~\ref{fig:architecture}.

\subsection{Perceptual contrast stretching - PCS}
\label{ssec:pcs}
PCS~\cite{b33} is a spectral processing technique that aims to improve the perceptual quality of a speech signal.
It leverages empirical observations that indicate different sensitivity levels in the human auditory system. 
PCS exploits this phenomenon by adjusting the magnitude spectrum of the signal according to each frequency band's perceived importance.
In this paper, we introduce PCS as an auxiliary step before the enhancement process, that is, a pre-processing step applied to both the noisy signal (input) and the clean signal (output). 

To apply PCS to a speech waveform, $\hat{x}(n)$, we first compute the STFT of $\hat{x}(n)$, which is then decomposed into magnitude $|STFT(\hat{x})|$ and phase $\angle STFT(\hat{x})$ components.
The magnitude is then compressed using a logarithmic function:

\begin{equation}
\Tilde{Y}(\omega, t) = \text{log}(|\text{STFT}(\hat{x})| + 1)
\end{equation}

The resulting compressed spectrum $\Tilde{Y}(\omega, t)$ is then stretched using the perceptual scores $W(\omega)$ derived by the Band Importance Function (BIF) proposed in~\cite{b33}:

\begin{equation}
\hat{Y}(\omega, t) = \Tilde{Y}(\omega, t) \cdot W(\omega)
\end{equation}

Where $\hat{Y}(\omega, t)$ is the perceptually stretched spectrum. 
It is then decompressed and combined with the original phase to be converted back to the time domain via iSTFT. 
As we show in the experimental section, this step has proven to be particularly effective in improving the perceptual quality of the enhanced signal estimated by SSL-based SE models.

\section{Dataset \& Experimental Setup}
\label{sec:experiments}

\subsection{Dataset}
\label{ssec:dataset}
The VoiceBank-DEMAND dataset~\cite{b36} was used to carry out our study.  VoiceBank-DEMAND  is a publicly available dataset that contains noisy speech recordings obtained by mixing clean speech from the VoiceBank data collection~\cite{b37}, and noise from the DEMAND dataset~\cite{b38}. The dataset covers a total of 30 distinct speakers, with 28 speakers used for training and the remaining 2 for testing.   The recording’s sampling frequency is 48 KHz, which were downsampled to 16 kHz as in \cite{b10}. Four types of signal-to-noise ratios (SNRs) were used to mix clean samples with noise samples in the dataset, [0, 5, 10, 15] dB for training and [2.5, 7.5, 12.5, 17.5] dB for testing. 
The total number of training utterances was 11572.  
The testing set included a total of 824 utterances. 
Training and test sets do not overlap in terms of speakers, noises, and signal-to-noise ratios.

\subsection{Experimental Setup}
\label{ssec:setup}
Following the standard practice \cite{b10}, all recordings were downsampled from 48 kHz to 16 kHz. Furthermore, the evaluation metrics used to assess our approach and compare with competitive SE solutions include: 
\begin{itemize}
    \item PESQ: Perceptual evaluation of speech quality, using the wide-band version recommended in ITU-T P.862.2 ~\cite{b39} (value ranges from –0.5 to 4.5).
    \item CSIG: Mean opinion score (MOS) prediction of the signal distortion attending only to the speech signal \cite{b40} (value ranges from 1 to 5).
    \item CBAK: MOS prediction of the intrusiveness of background noise \cite{b40} (value ranges from 1 to 5).
    \item COVL: MOS prediction of the overall effect \cite{b40} (value ranges  from 1 to 5).
    \item Short-time objective intelligibility (STOI) \cite{b41} (value ranges from 0 to 1)
\end{itemize}

In our experiments, to achieve temporal alignment between the outputs of the SSL model and the STFT, we configured the STFT parameters as follows: the number of FFT points was set to 400, the hop length was 160, and the window length was 400.
The SSL model architecture consists of a CNN front-end that extracts feature representations from the raw waveform input, followed by a series of transformer layers that take the CNN features as input.
In this context, to synchronize the SSL and STFT representations along the temporal dimension, we adapted the stride of the final convolutional layer in the CNN to be 1. Differently from previous works~\cite{b26, b27} that align representations via frame duplication, this adaptation provides a one-to-one correspondence between the SSL and STFT frames~\cite{b42}. During training, the maximum input length was set to 10 seconds, applying padding when sequences were shorter than this value. 
When running inference on test input signals, the model preserves the original sequence length.
The features dimension of the SSL model output is 1024. As input to the SE head, we concatenated  1024-dim SSL and 201-dim STFT features, yielding a 1225-dim input representation.

To guarantee a fair comparison among the various proposed model variants, we maintain consistent training and testing setups across all experiments. All models are trained on a single NVidia A100 GPU and the overall architecture is trained for 50 epochs with the Adam optimizer, using a learning rate of $10^{-4}$. We leverage the pre-trained WavLM model
\footnote{The pre-trained model is available at \texttt{https://huggingface.co/microsoft/wavlm-large}} as the backbone for our experiments.

Our investigation also covers the evaluation of different SE heads, 
and PCS on the overall performance\footnote{https://github.com/salman18376/SE-SSL}.

\section{Experimental Results}
\label{sec:results}

\subsection{SSL based results}
\label{subsec:SSLbasedresults}









\begin{table}[]
\centering
\caption{Comparison between proposed CS-WavLM SE model and  best SE solutions tested on the VoiceBank+DEMAND dataset. SSL-based solutions are indicated by a $^\star$.}
\label{tab:main_results}
\resizebox{\columnwidth}{!}{%
\begin{tabular}{llllll}
\hline
Model                     & PESQ          & CSIG          & CBAK          & COVL          & STOI          \\ \hline
CMGAN \cite{b11}          & 3.41          & 4.63          & 3.94          & 4.12          & \textbf{0.96} \\
TridentSE \cite{b43}      & 3.47          & 4.70          & 3.81          & 4.10          & \textbf{0.96} \\
MP-SENet \cite{b21}       & 3.50          & 4.73          & \textbf{3.95} & 4.22          & \textbf{0.96} \\ \hline
$^{*}$BSSE \cite{b26}     & 3.20          & 4.53          & 3.78          & 4.04          & \textbf{0.96} \\
$^{*}$SSF-CVAE \cite{b44} & 3.04          & 4.38          & 2.91          & 3.72          & 0.95          \\
$^{*}$CS-WavLM            & 3.29          & 4.64          & 3.80          & 4.05          & \textbf{0.96} \\ \hline
$^{*}$PCS-BSSE            & 3.46          & \textbf{4.75} & 3.49          & 4.20          & 0.95          \\
$^{*}$PCS-CS-WavLM        & \textbf{3.54} & \textbf{4.75} & 3.54          & \textbf{4.25} & \textbf{0.96} \\ \hline
\end{tabular}%
}
\end{table}

Table~\ref{tab:main_results} shows the result attained by our method, namely PCS ConSistent WavLM (PCS-CS-WavLM), alongside those of other recently proposed SE methods. 
In particular, state-of-the-art (SOTA) SE solutions on the VoiceBank-DEMAND are listed in the upper part of the table; whereas, SSL-based methods are given in the central part of the table, where BSSE refers to the SSL-based technique presented in \cite{b26} and  SSF-CVAE refers to \cite{b44}. 
A comparison between BSSE, SSF-CVAE, and our CS-WavLM, in the fourth, fifth and sixth row, respectively,  suggests that our approach notably outperforms previously proposed best SSL-based solutions \cite{b26,b44} assessed on the same task. 
Furthermore, comparing the first three rows with the sixth row, we observe that CS-WavLM reduced the gap with SOTA SE approaches. 
When applying PCS pre-processing to both noisy and target waveforms before training, PCS-CS-WavLM achieves the best results in terms of PESQ, CIG, COVL, and STOI.
Interestingly, PCS significantly improves the perceptual quality of the enhanced signal (e.g., +0.25 in PESQ), but it also causes a slight reduction of CBAK. 
For completeness, we also include the results obtained by applying PCS pre-processing to train the BSSE model, i.e., PCS-BSSE in the 7th row.

\subsection{Ablation studies}
\label{subsec:ablation}

To analyze the impact of the different components of the proposed SSL-based SE solution, we carry out specific ablation studies concerning (i) the architectural choices for the SE head, 
and (ii) the effect of PCS processing.

\vspace{1.0mm}
\noindent
\textbf{SE head:} The top part of Table~\ref{tab:ablation_studies} is concerned with the performance attained by the proposed SE solution using different heads. In particular,  we compare  Conformer-,  BiLSTM-, and Transformer-based heads. The Conformer-based head is the one already used in the PCS-CS-WavLM system and reported in the last row in Table~\ref{tab:main_results}. 

The BiLSTM head shows a  drop in performance compared to the Tranformer head, potentially due to its inability to effectively model long-term dependencies in the input sequence compared to the self-attention used in Transformer. 
The Conformer-based head used in PCS-CS-WavLM attains the best results, which could be due to the combination local and global information.  


\vspace{1.0mm}
\noindent
\textbf{PCS:}
To explore the impact of PCS processing  on both input and target signals at a training time, we analyze various configurations. In particular, we compare configurations where PCS pre-processing is applied to both input and target signals, only to input signals, or only to target signals. It should be noted that when PCS is used during testing, PCS is only applied to the input noisy waveforms.
Interestingly, the combined use of PCS on both input and target signals (i.e., PCS-CS-WavLM) consistently outperforms other configurations, attaining the highest values across all metrics.
Applying PCS pre-processing only on the input signals yields better results compared to performing PCS on the  target signals only. Nonetheless, the best results can be attained pre-processing both input and target signals, as it can be verified by comparing the proposed  PCS-CS-WavLM models (last row) with the first two rows in the PCS section in Table~\ref{tab:ablation_studies}. We argue that PCS  limited to either the input or output signals lead to a mismatch in what the model learns, as one of the two signals does not undergo the same pre-processing.

\begin{table}[]
\centering
\caption{Effect on SSL-based SE performance of varying components and processes.}
\label{tab:ablation_studies}
\resizebox{\columnwidth}{!}{%
\begin{tabular}{llllll}
\hline
Model          & PESQ          & CSIG          & CBAK          & COVL          & STOI          \\ \hline
\multicolumn{6}{c}{SE head}                                                                    \\ \hline
BiLSTM         & 3.47          & 4.70          & 3.50          & 4.18          & 0.95          \\
Transformer    & 3.52          & 4.73          & 3.53          & 4.23          & 0.95          \\ \hline
\multicolumn{6}{c}{PCS pre-processing}                                                         \\ \hline
w/o PCS input  & 3.17          & 4.16          & 3.27          & 3.71          & 0.94          \\
w/o PCS target & 3.31          & 4.65          & \textbf{3.70} & 4.06          & 0.96          \\ \hline
PCS-CS-WavLM   & \textbf{3.54} & \textbf{4.75} & 3.54          & \textbf{4.25} & \textbf{0.96} \\ \hline
\end{tabular}%
}
\end{table}


\section{Conclusion}
\label{sec:conclusion}

We have proposed to leverage perceptual contrast stretching, and a consistency-preserving loss in order to boost SSL-based speech enhancement solutions with the goal of closing the gap with state-of-the-art solutions. Our approach, referred to as PCS-CS-WavLM, uses a pre-trained WavLM backbone with a Conformer-based SE head, and three losses, namely a waveform-based loss, and two magnitude-based losses. For one of the two magnitude-based losses, we followed the consistency-preserving idea outlined in~\cite{b12}, which was useful to improve the overall PESQ. Our approach has been assessed against the  VoiceBank+DEMAND task, reporting competitive results when combined with PCS.  Furthermore, our ablation studies have demonstrated (i) the benefits of the Conformer architecture for improving SE quality, and 
(ii) the importance of performing PCS on both input and target signals. 
In sum, our experimental evidence supported our claim to improve SSL-based SE systems and make a step forward to close the gap with SOTA SE solutions.
In future work, we intend to explore self-supervised pre-training objectives that may be better tailored to the enhancement task.


\vspace{12pt}
\color{red}

\end{document}